\documentclass[11pt,a4paper]{article}
\pdfoutput=1
\usepackage{jheppub}
\usepackage{tikz}
\usepackage{subcaption}
\DeclareMathOperator{\Tr}{Tr}

\newcommand{\ri}{\mathrm{i}}

\newcommand{\qu}{\frac{1}{4}}

\newcommand{\si}{\sigma}

\newcommand{\del}{\partial}

\newcommand{\bra}{\langle}
\newcommand{\ket}{\rangle}
\newcommand{\ketc}{\rangle_{c}}
\newcommand{\la}{\lambda}

\newcommand{\bt}{\beta}

\newcommand{\al}{\alpha}

\newcommand{\rt}[1]{\sqrt{#1}}
\newcommand{\cZ}{\mathcal{Z}}
\newcommand{\cE}{\mathcal{E}}
\newcommand{\cO}{\mathcal{O}}

\newcommand{\cH}{\mathcal{H}}

\begin{document}

\title{Replica symmetry breaking in random matrix model: a toy model of wormhole networks}

\author{Kazumi Okuyama}

\affiliation{Department of Physics, 
Shinshu University,
3-1-1 Asahi, Matsumoto 390-8621, Japan}

\emailAdd{kazumi@azusa.shinshu-u.ac.jp}

\abstract{We study the replica symmetry breaking (RSB) in
the Gaussian Unitary Ensemble (GUE)
random matrix model.
We find that the RSB
occurs at the transition temperature 
$T_{\text{RSB}}\sim N^{-2/3}$
in the large $N$ limit.
We argue that this  transition
originates from a landscape of almost degenerate 
local minima of free energy labeled by Young diagrams.
We also discuss a possible implication of our findings
for the multiply-coupled Sachdev-Ye-Kitaev models and their holographic 
dual Jackiw-Teitelboim gravity from the viewpoint of ER=EPR conjecture.}

\maketitle

\section{Introduction}
\label{sec:intro}

The replica symmetry breaking (RSB)
in disordered system is an important concept in the
study of glass transition.
In particular, the spin-glass model of Sherrington and Kirkpatrick 
\cite{SKmodel} was solved by Parisi \cite{parisi1,parisi2}
based on the RSB ansatz
(see \cite{Denef:2011ee} for a review).

In the disordered system a natural quantity to compute 
is the quenched average of the free energy $\bra \log Z(\bt)\ket$,
where $Z(\bt)=\Tr e^{-\bt H}$ is the partition function and 
$\bt$ is the inverse of temperature $T$
\begin{equation}
\begin{aligned}
 \bt=\frac{1}{T}.
\end{aligned} 
\end{equation}
The average $\bra \cdots\ket$ in the disordered system
is defined by integrating over the distribution
of random couplings in the Hamiltonian $H$. 
The replica method is commonly used to compute 
the quenched average $\bra \log Z(\bt)\ket$ by
preparing $n$ copies of the system (replicas) 
and taking the limit $n\to0$ at the end
of calculation
\begin{equation}
\begin{aligned}
\bra \log Z(\bt)\ket= \lim_{n\to0}\frac{\del}{\del n}\bra Z(\bt)^n\ket.
\end{aligned} 
\label{eq:n-lim}
\end{equation}
RSB is characterized by the condition
that the quenched average and the annealed average 
of $n$ replicas do not agree
\begin{equation}
\begin{aligned}
 \bra Z(\bt)^n\ket\ne \bra Z(\bt)\ket^n.
\end{aligned} 
\label{eq:cond-RSB}
\end{equation}

In this paper, we consider the RSB in the Gaussian Unitary Ensemble (GUE)
random matrix model,
where the random couplings in the Hamiltonian are modeled
by treating the Hamiltonian $H$ itself as a random  
$N\times N$ hermitian matrix
with the Gaussian distribution
\begin{equation}
\begin{aligned}
 \bra \cdots \ket=\frac{\int dH e^{-\frac{N}{2}\Tr H^2}(\cdots)}{\int dH e^{-\frac{N}{2}\Tr H^2}}.
\end{aligned} 
\label{eq:Gauss}
\end{equation}
Note that the RSB
in the GUE random matrix model has been considered in \cite{kamenev}, where
the correlator of resolvent $\Tr (E-H)^{-1}$
was studied by the replica method.
In this paper we consider the
correlator of $Z(\bt)=\Tr e^{-\bt H}$ and
study the RSB characterized by \eqref{eq:cond-RSB}.\footnote{The
correlator $\bra Z(\bt)^n\ket$ in random matrix model
has been studied for a long time
in the context of  2d quantum
gravity under the name of ``loop operators'' \cite{Ginsparg:1993is}.}
We will not consider the limit $n\to 0$ in \eqref{eq:n-lim}; we
study $\bra Z(\bt)^n\ket$ with positive integer $n$ in its own right.

We find that in the large $N$ limit $\bra Z(\bt)^n\ket$ in the GUE random matrix
model indeed exhibits the RSB at the transition temperature
$T_{\text{RSB}}$
\begin{equation}
\bra Z(\bt)^n\ket\approx\left\{
\begin{aligned}
 &\bra Z(\bt)\ket^n, &\quad  &(T>T_{\text{RSB}}),\\
&\bra Z(\bt)^n\ket_c, &\quad &(T<T_{\text{RSB}}),
\end{aligned} 
\right.
\label{eq:RSB}
\end{equation}
where the subscript $c$ in $\bra Z(\bt)^n\ket_c$
denotes the connected part of the correlator.
A similar behavior was observed in the so-called random-energy 
model, where each energy levels are assumed to be Gaussian distributed but
the correlations between different energy levels 
are neglected \cite{Derrida}\footnote{
In the high temperature regime $T>T_{\text{RSB}}$,
taking the limit in \eqref{eq:n-lim} we find
$\bra \log Z(\bt)\ket\approx \log\bra Z(\bt)\ket$.
On the other hand,
the $n\to0$ limit in the low temperature regime
$T<T_{\text{RSB}}$ is rather subtle and the analytic continuation to $n=0$ 
is not straightforward \cite{Derrida}. As emphasized in \cite{Derrida},
the correct free energy is not  reproduced by simply
taking the 
$n\to0$ limit of $\bra Z(\bt)^n\ket_c$ in \eqref{eq:RSB}.
}.
It turns out that $T_{\text{RSB}}$
for the GUE matrix model scales as
\begin{equation}
\begin{aligned}
 T_{\text{RSB}}\sim N^{-2/3}.
\end{aligned} 
\label{eq:glass-T}
\end{equation}
We will argue that the transition at $T=T_{\text{RSB}}$
originates from the existence of
a landscape of almost degenerate local minima of free energy.
This landscape of local minima arises
as follows: When we expand $\bra Z(\bt)^n\ket$
in terms of the connected correlators, there appear many terms labeled by Young diagrams.
Each term of this expansion can be thought of as a local minimum of free energy.

We will also discuss a possible implication of our findings for
the Sachdev-Ye-Kitaev (SYK) model \cite{kitaev,Maldacena:2016hyu}
and its holographic dual Jackiw-Teitelboim (JT) gravity \cite{Jackiw:1984je,Teitelboim:1983ux,Almheiri:2014cka}.
We examine the idea that the connected part of correlator
$\bra Z(\bt)^n\ket_c$
corresponds to a dual spacetime geometry where
the $n$ boundaries are connected by an $n$-pronged Euclidean wormhole.
We speculate that around the RSB transition temperature the dual gravity side
does not corresponds to a classical geometry, but 
a superposition of random wormhole networks.

This paper is organized as follows.
In section \ref{sec:exact}, we first review the exact result of $\bra Z(\bt)^n\ket$
at finite $N$ in the GUE random matrix model.
In section \ref{sec:RSB}, we study the RSB of $\bra Z(\bt)^n\ket$ in the GUE random matrix model
using the exact result at finite $N$.
In section \ref{sec:SYK}, we discuss possible implications of our findings
in the GUE matrix model for the multiply-coupled SYK models
and their dual JT gravity. 

\section{Review of the exact results of GUE random matrix model}
\label{sec:exact}

In this section, we review the known results of the correlation function
$\bra Z(\bt)^n\ket$  in the GUE random matrix model
\eqref{eq:Gauss}.
As pointed out in \cite{delCampo:2017bzr,Okuyama:2018yep},
this quantity is formally equivalent to the expectation value
of the 1/2 BPS Wilson loops in 4d $\mathcal{N}=4$ $U(N)$ super Yang-Mills (SYM) theory
under the identification
\begin{equation}
\begin{aligned}
 \rt{\la}\leftrightarrow 2\bt,
\end{aligned} 
\label{eq:la-dic}
\end{equation}
where $\la$ is the 't Hooft coupling of $\mathcal{N}=4$ SYM.
Thus we can immediately find the 
exact finite $N$ expression of  $\bra Z(\bt)^n\ket$ by borrowing
the known result of $\mathcal{N}=4$ SYM.

We can naturally decompose $\bra Z(\bt)^n\ket$
into the connected components $\bra Z(\bt)^j\ket_c~(j=1,\cdots,n)$.
The relation between $\bra Z(\bt)^n\ket$
and the connected part $\bra Z(\bt)^j\ket_c$
is compactly expressed in terms of the generating function, as usual
\begin{equation}
\begin{aligned}
\bigl\bra e^{\al Z(\bt)}\bigr\ket=\sum_{n=0}^\infty \frac{\al^n}{n!}\bra Z(\bt)^n\ket=\exp\Biggl[
\sum_{j=1}^\infty \frac{\al^j}{j!}\bra Z(\bt)^j\ketc\Biggr],
\end{aligned} 
\end{equation}
where $\al$ is a formal expansion parameter.
For instance, $\bra Z(\bt)^n\ket$ with $n=2,3,4$ are expanded as
\begin{equation}
\begin{aligned}
 \bra Z(\bt)^2\ket&= \bra Z(\bt)\ket^2+\bra Z(\bt)^2\ket_c,\\
  \bra Z(\bt)^3\ket&= \bra Z(\bt)\ket^3+3\bra Z(\bt)\ket\bra Z(\bt)^2\ket_c+
\bra Z(\bt)^3\ket_c,\\
\bra Z(\bt)^4\ket&= \bra Z(\bt)\ket^4+6\bra Z(\bt)\ket^2\bra Z(\bt)^2\ket_c+
3\bra Z(\bt)^2\ket_c^2+4\bra Z(\bt)\ket\bra Z(\bt)^3\ket_c
+\bra Z(\bt)^4\ket_c.
\end{aligned} 
\label{eq:zconn-exp}
\end{equation}
In general, the expansion of $\bra Z(\bt)^n\ket$ is characterized by the partition of $n$
labeled by a Young diagram $Y$
\begin{equation}
\begin{aligned}
 Y=[1^{k_1}2^{k_2}\cdots n^{k_n}],\qquad |Y|=\sum_{j=1}^n jk_j=n,
\end{aligned} 
\label{eq:defY}
\end{equation}
and \eqref{eq:zconn-exp} for general $n$ becomes
\begin{equation}
\begin{aligned}
 \bra Z(\bt)^n\ket=\sum_{|Y|=n}\cZ_Y(\bt),
\end{aligned} 
\label{eq:Y-sum}
\end{equation}
where $\cZ_Y(\bt)$ is given by\footnote{A similar expansion has also appeared in \cite{Derrida}.}
\begin{equation}
\begin{aligned}
 \cZ_Y(\bt)=n!\prod_{j=1}^{n}\frac{1}{k_j!}\left[\frac{\bra Z(\bt)^j\ket_c}{j!}\right]^{k_j}.
\end{aligned} 
\end{equation}
For example, $[1^n]$ and $[n]$ correspond to
the totally disconnected and the totally connected part, respectively
\begin{equation}
\begin{aligned}
 \cZ_{[1^n]}(\bt)=\bra Z(\bt)\ket^n,\qquad
\cZ_{[n]}(\bt)=\bra Z(\bt)^n\ket_c.
\end{aligned} 
\end{equation}

As shown in \cite{Okuyama:2018aij},
we can write down the exact finite $N$ result of $\bra Z(\bt)^n\ket_c$ 
in terms of the $N\times N$ matrix $A(\bt)$
\begin{equation}
\begin{aligned}
 A(\bt)_{i,j}=e^{\frac{\bt^2}{2N}}\rt{\frac{i!}{j!}}\left(\frac{\bt}{\rt{N}}\right)^{j-i}
L_{i}^{j-i}\left(-\frac{\bt^2}{N}\right),\qquad
(i,j=0,\cdots, N-1),
\end{aligned} 
\end{equation}
where $L_n^\al(x)$ denotes the associated Laguerre polynomial.
The one-point function $\bra Z(\bt)\ket$ is simply given by
the trace of $A(\bt)$
\begin{equation}
\begin{aligned}
\bra Z(\bt)\ket=\Tr A(\bt)=e^{\frac{\bt^2}{2N}} L_{N-1}^1\left(-\frac{\bt^2}{N}\right).
\end{aligned} 
\label{eq:z-L}
\end{equation}
Similarly, the connected part $\bra Z(\bt)^n\ket_c$
is given by some combination of $A(\bt)$ \cite{Okuyama:2018aij}
\begin{equation}
\begin{aligned}
 \bra Z(\bt)^n\ket_c
=\oint \prod_{i=1}^n\frac{dz_i}{2\pi\ri z_i^2}
\Tr\log\left[\sum_{m=0}^n \sum_{i_1<\cdots <i_m}z_{i_1}\cdots z_{i_m}A(m\bt)\right].
\end{aligned} 
\end{equation}
For instance, $\bra Z(\bt)^2\ket_c$ is given by
\begin{equation}
\begin{aligned}
 \bra Z(\bt)^2\ket_c
&=\oint \frac{dz_1}{2\pi\ri z_1^2}\oint \frac{dz_2}{2\pi\ri z_2^2}
\Tr\log\Bigl[1+z_1A(\bt)+z_2A(\bt)+z_1z_2A(2\bt)\Bigr]\\
&=\Tr\Bigl[A(2\bt)-A(\bt)^2\Bigr].
\end{aligned} 
\label{eq:z2-A}
\end{equation}
In a similar manner, one can write down $\bra Z(\bt)^3\ket_c$ and
$\bra Z(\bt)^4\ket_c$ as
\begin{equation}
\begin{aligned}
\bra Z(\bt)^3\ket_c&=\Tr\Bigl[A(3\bt)-3A(\bt)A(2\bt)+2A(\bt)^3\Bigr],\\
\bra Z(\bt)^4\ket_c&=\Tr\Bigl[A(4\bt)-4A(\bt)A(3\bt)-3A(2\bt)^2+12A(\bt)^2A(2\bt)-6A(\bt)^4\Bigr].
\end{aligned} 
\label{eq:z34-A}
\end{equation}
Using these exact results \eqref{eq:z2-A} and \eqref{eq:z34-A}, 
in the next section
we will numerically study the behavior of $\bra Z(\bt)^n\ket~(n=2,3,4)$ 
in \eqref{eq:zconn-exp}.

Next, let us briefly summarize the known results of
$\bra Z(\bt)^n\ket$ in the large $N$ limit.
In the large $N$ limit, the eigenvalues
of $H$ with the Gaussian measure \eqref{eq:Gauss}
are distributed along the cut $[E_0,-E_0]$
with the eigenvalue density $\rho(E)$ obeying the Wigner semi-circle law
\begin{equation}
\begin{aligned}
 \rho(E)=\frac{2}{\pi E_0^2}\rt{E_0^2-E^2}.
\end{aligned} 
\label{eq:wigner}
\end{equation}
In our normalization of the measure  \eqref{eq:Gauss},
the ``ground state energy'' $E_0$ is given by
\begin{equation}
\begin{aligned}
 E_0=-2.
\end{aligned} 
\label{eq:E0}
\end{equation}

The eigenvalue density \eqref{eq:wigner}
is a property of a single eigenvalue.
One can consider the correlation of $n$ eigenvalues
characterized by the correlation function $\rho^{(n)}(E_1,\cdots, E_n)$.
It turns out that $\rho^{(n)}$ is written in terms of the kernel $K$ 
(see e.g. \cite{Mehta})
\begin{equation}
\begin{aligned}
 K(E,E')=\sum_{\ell=0}^{N-1} \psi_\ell(E)\psi_\ell(E'),
\end{aligned} 
\end{equation}
where $\psi_\ell(E)$ is the wavefunction associated with
the measure \eqref{eq:Gauss}, i.e. 
the wavefunction of harmonic oscillator in this case
\begin{equation}
\begin{aligned}
 \psi_\ell(E)=\left(\frac{N}{2\pi}\right)^\qu
\frac{1}{\rt{2^\ell\ell!}}H_\ell\left(\rt{\frac{N}{2}}E\right)e^{-\frac{NE^2}{4}}.
\end{aligned} 
\end{equation}
Here $H_\ell(x)$ denotes the Hermite polynomial of order $\ell$.
Note that 
the eigenvalue density is given by the diagonal part of the kernel $K$
\begin{equation}
\begin{aligned}
 \rho(E)=\frac{1}{N}K(E,E).
\end{aligned} 
\end{equation}
Now $\bra Z(\bt)^n\ket$ is written in terms of $K$ as
\begin{equation}
\begin{aligned}
 \bra Z(\bt)^n\ket&=\int \prod_{i=1}^n dE_i e^{-\bt E_i}
\det\Bigl[K(E_k,E_l)\Bigr]_{k,l=1,\cdots,n}\\
&=\sum_{\si\in S_n}(-1)^\si \int \prod_{i=1}^n dE_i e^{-\bt E_i} K(E_i,E_{\si(i)}).
\end{aligned} 
\label{eq:det-K}
\end{equation}
One can show that the decomposition \eqref{eq:Y-sum}
corresponds to the sum over the 
conjugacy class of $\si\in S_n$ in \eqref{eq:det-K}.
The totally connected part $\cZ_{[n]}(\bt)=\bra Z(\bt)^n\ket_c$
comes from the cyclic permutation with length $n$
\begin{equation}
\begin{aligned}
 \bra Z(\bt)^n\ket_c=(-1)^{n-1}\sum_{a=1}^{n-1}\int \prod_{i=1}^n
dE_i e^{-\bt E_i}
K(E_i,E_{i+a})
\end{aligned} 
\label{eq:rho-c}
\end{equation}
where the subscript of $E_i$ is defined modulo $n$ ($E_{i+n}\equiv E_i$).

The leading order behavior
of $\bra Z(\bt)^n\ket_c$ in the large $N$ expansion 
can be easily found by borrowing
the known results of $\mathcal{N}=4$ SYM in
\cite{Erickson:2000af,Akemann:2001st,Giombi:2009ms}
via the dictionary \eqref{eq:la-dic}
\begin{equation}
\begin{aligned}
 \bra Z(\bt)\ket&=\frac{N}{\bt}I_1(2\bt),\\
 \bra Z(\bt)^2\ket_c&=\bt I_0(2\bt)I_1(2\bt),\\
\bra Z(\bt)^3\ket_c&=\frac{\bt^3}{N}
\Bigl[I_0(2\bt)^3+3I_0(2\bt)^2I_1(2\bt)\Bigr],
\end{aligned} 
\label{eq:z-bessel}
\end{equation}
where $I_\nu(z)$ denotes the modified Bessel function of the 
first kind. 
Note that in the large $N$ limit $\bra Z(\bt)^n\ket_c$ scales as
\begin{equation}
\begin{aligned}
 \bra Z(\bt)^n\ket_c\sim N^\chi,
\end{aligned} 
\end{equation}
where $\chi$ is the Euler number of a sphere with $n$ punctures
\begin{equation}
\begin{aligned}
 \chi=2-n.
\end{aligned}
\label{eq:chi-def} 
\end{equation}
\section{RSB in the GUE random matrix model}
\label{sec:RSB}
In this section, we will study the RSB in 
the GUE random matrix model using the exact result reviewed
in the previous section \ref{sec:exact}.

\subsection{Numerical analysis of RSB}
\label{sec:num}

\begin{figure}[thb]
\centering
\subcaptionbox{$R_2(\bt)$\label{sfig:R2}}{\includegraphics[width=4.5cm]{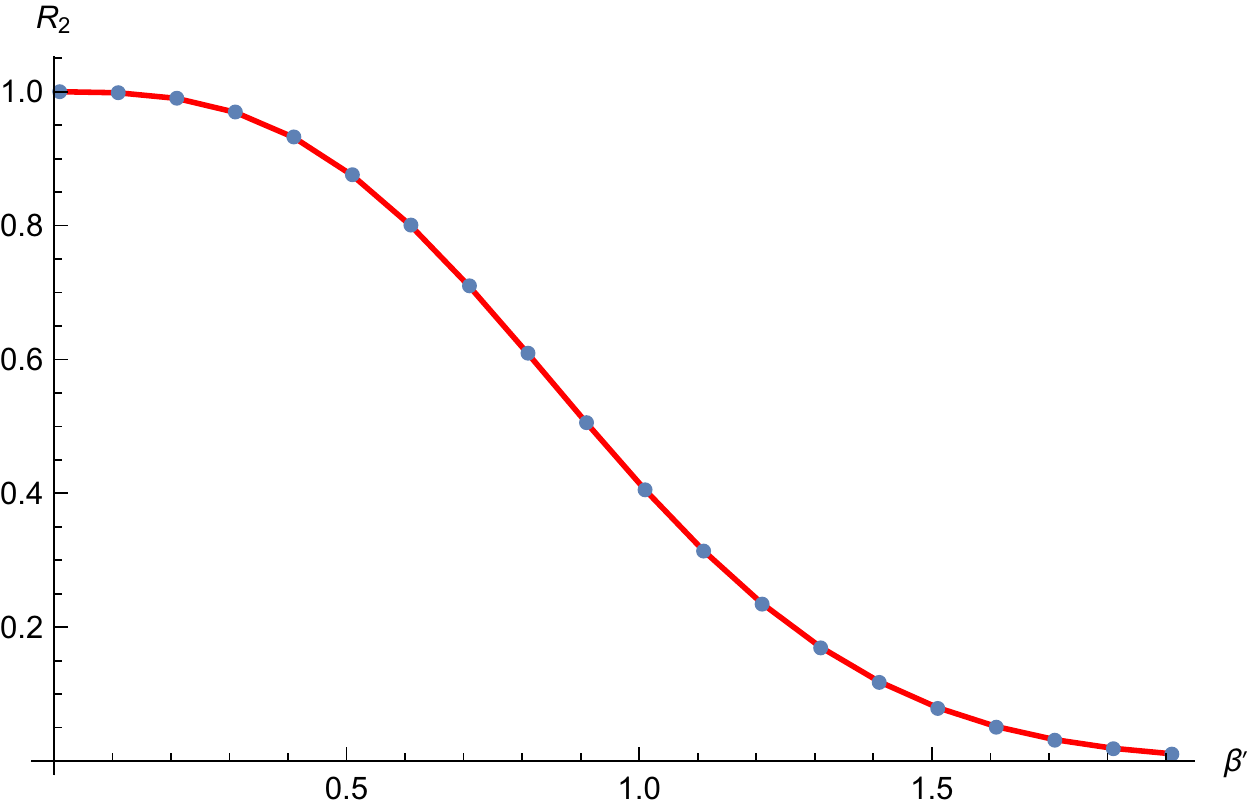}}
\hskip2mm
\subcaptionbox{$R_3(\bt)$\label{sfig:R3}}{\includegraphics[width=4.5cm]{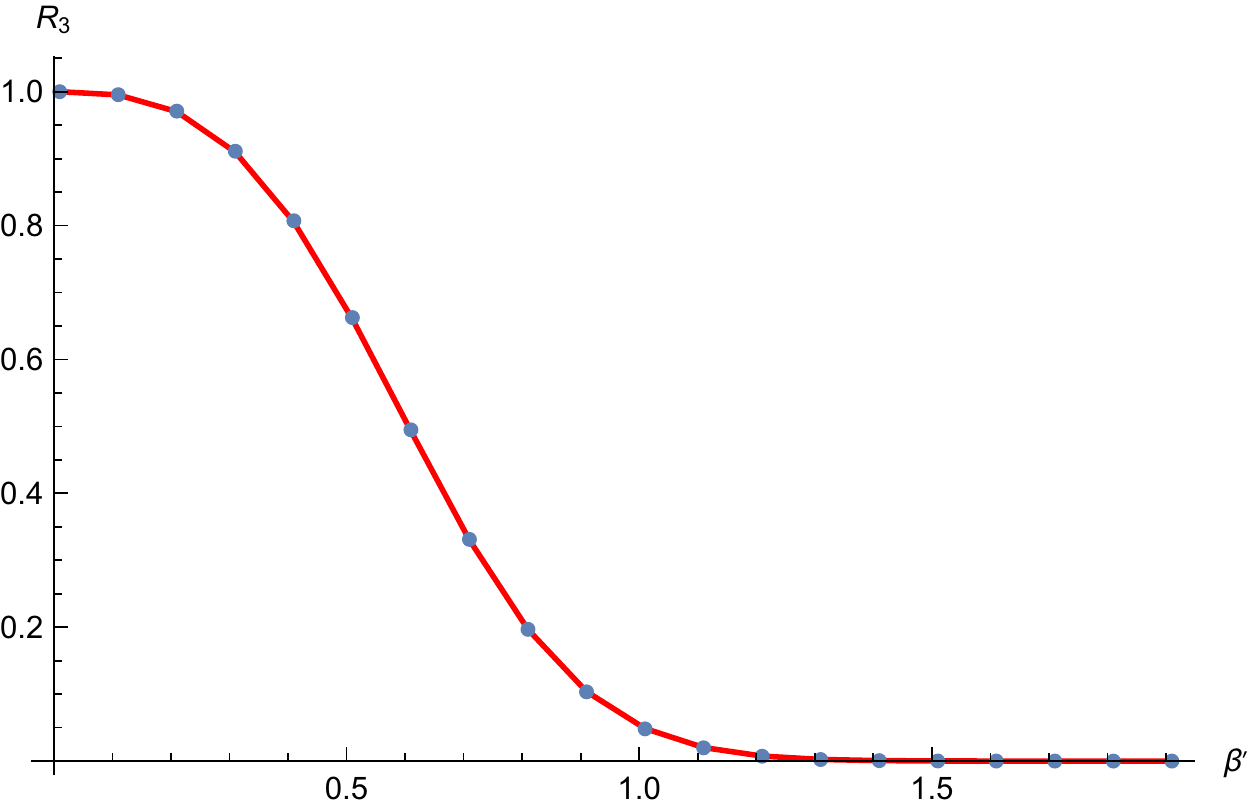}}
\hskip2mm
\subcaptionbox{$R_4(\bt)$\label{sfig:R4}}{\includegraphics[width=4.5cm]{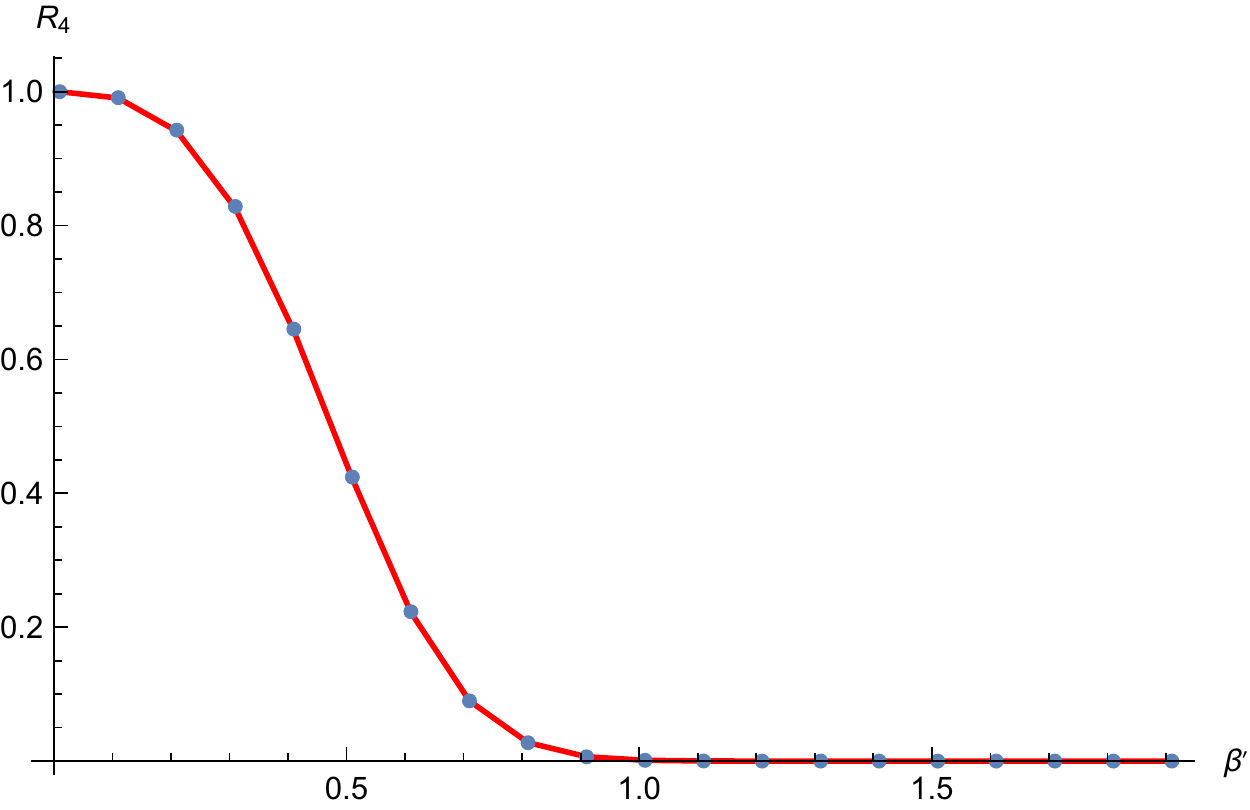}}
  \caption{Plot of $R_n(\bt)$
for \subref{sfig:R2} $n=2$, \subref{sfig:R3} $n=3$, and \subref{sfig:R4} $n=4$.
Note that the horizontal axis is $\bt'=\bt/N^{2/3}$.
The red curve is the plot for $N=60$, while the blue dots represent 
the value of $R_n(\bt)$ for $N=100$.
}
  \label{fig:Rn}
\end{figure}

To see the RSB in \eqref{eq:RSB},
let us consider the ratio
between the annealed
average $\bra Z(\bt)\ket^n$ and the quenched average $\bra Z(\bt)^n\ket$
\begin{equation}
\begin{aligned}
 R_n(\bt)
=\frac{\bra Z(\bt)\ket^n}{\bra Z(\bt)^n\ket}.
\end{aligned} 
\label{eq:Rn-def}
\end{equation}
We can numerically 
study the behavior of $R_n(\bt)~(n=2,3,4)$ as a function of $\bt$
using the exact result of $\bra Z(\bt)\ket$ in \eqref{eq:z-L}
and $\bra Z(\bt)^n\ket_c~(n=2,3,4)$  in \eqref{eq:z2-A}
and \eqref{eq:z34-A},
together with the relation \eqref{eq:zconn-exp}.
It turns out that $R_n(\bt)$ is conveniently
described by the scaling variable $\bt'$ defined by
\begin{equation}
\begin{aligned}
 \bt'=\frac{\bt}{N^{2/3}}.
\end{aligned} 
\label{eq:bt'}
\end{equation}
In Figure \ref{fig:Rn}, we plot $R_n(\bt)~(n=2,3,4)$ as a function of $\bt'$
for $N=60$ (red curves) and $N=100$ (blue dots).
As we can see from Figure \ref{fig:Rn},
the plots of $R_n(\bt)$ for $N=60$ and $N=100$ are almost identical
in the region $\bt'\lesssim 2$.
This justifies a posteriori our definition of the scaling variable $\bt'$ in \eqref{eq:bt'}.
One can see from Figure \ref{fig:Rn}
that $R_n(\bt)$ decays to zero about $\bt'\sim 1$.
This indicates 
that the replica symmetry is broken in the low temperature regime $\bt'\gtrsim 1$, 
and the critical value of $\bt$ is roughly given by
\begin{equation}
\begin{aligned}
 \bt_{\text{RSB}}\sim N^{2/3},
\end{aligned} 
\label{eq:bt-glass}
\end{equation}
which corresponds to the temperature $T_{\text{RSB}}$ in \eqref{eq:glass-T}.
In the next subsection, we consider an analytic explanation of
the $N$-dependence of the transition temperature in \eqref{eq:bt-glass}.

For $n=2$, the decay of $R_2(\bt)$ means that in the low temperature region $T<T_{\text{RSB}}$
the connected part $\bra Z(\bt)^2\ket_c$ becomes dominant, and
around the transition temperature
$T\sim T_{\text{RSB}}$ 
it takes place the exchange of dominance between the connected part 
$\bra Z(\bt)^2\ket_c$ and the
disconnected part $\bra Z(\bt)\ket^2$. 
However, for $n\geq3$ what is happening around $T\sim T_{\text{RSB}}$
is more involved since $\bra Z(\bt)^n\ket$ has many terms
\eqref{eq:Y-sum} when written in terms of the combination of connected correlators. 
We will study the detailed structure for $n\geq3$ in section \ref{sec:glass}.

\subsection{Analytic estimate of the transition temperature}
\label{sec:estimate}
In this subsection,
we will estimate the transition temperature using
the large $N$ result of $\bra Z(\bt)^n\ket_c$ in \eqref{eq:z-bessel}.
From the asymptotic behavior of the modified Bessel function
\begin{equation}
\begin{aligned}
 I_\nu(z)\sim \frac{e^z}{\rt{2\pi z}},\qquad (z\gg1),
\end{aligned} 
\label{eq:I-asymp}
\end{equation}
one can see that 
in the regime $\bt\gg1$, $\bra Z(\bt)^n\ket_c$ in \eqref{eq:z-bessel}
behaves as
\begin{equation}
\begin{aligned}
 \bra Z(\bt)^n\ket_c\sim\left(\frac{N}{\bt^{3/2}}\right)^{2-n}e^{2n\bt}.
\end{aligned} 
\label{eq:z-eff}
\end{equation}
As discussed in \cite{Drukker:2000rr}, the combination
$\bt^{3/2}/N$ appearing in \eqref{eq:z-eff}
has a natural interpretation as the effective genus counting parameter
\begin{equation}
\begin{aligned}
 g_{\text{eff}}=\frac{\bt^{3/2}}{N}.
\end{aligned} 
\label{eq:g-eff}
\end{equation}
Indeed, 
the large $N$ expansion of the exact finite $N$
result \eqref{eq:z-L}, \eqref{eq:z2-A}
and \eqref{eq:z34-A}
beyond the leading order \eqref{eq:z-bessel}
are organized as the genus expansion with 
the effective string coupling in \eqref{eq:g-eff} 
\cite{Drukker:2000rr}.
In terms of this effective string coupling $g_{\text{eff}}$,
\eqref{eq:z-eff} is written as
\begin{equation}
\begin{aligned}
 \bra Z(\bt)^n\ket_c\sim g_{\text{eff}}^{-\chi}e^{-n\bt E_0}
\end{aligned} 
\label{eq:geff-chi}
\end{equation}
where $\chi$ is defined in \eqref{eq:chi-def} and
$E_0=-2$ \eqref{eq:E0}.
\eqref{eq:geff-chi} should be compared with the disconnected part
\begin{equation}
\begin{aligned}
 \bra Z(\bt)\ket^n\sim \Bigl(g_{\text{eff}}^{-1}e^{-\bt E_0}\Bigr)^n
=g_{\text{eff}}^{-n}e^{-n\bt E_0}.
\end{aligned} 
\label{eq:disc-sim}
\end{equation}
The connected part \eqref{eq:geff-chi} and
the disconnected part \eqref{eq:disc-sim} become comparable when
the effective string coupling becomes of order $1$
\begin{equation}
\begin{aligned}
 g_{\text{eff}}\sim 1.
\end{aligned} 
\label{eq:geff-1}
\end{equation}
From \eqref{eq:g-eff}, this condition explains the $N$-dependence of the
transition temperature $\bt_{\text{RSB}}$
in \eqref{eq:bt-glass}.

The scaling \eqref{eq:z-eff}
for $n=1$ is easily understood from the behavior
of the eigenvalue density near the ground state $E\sim E_0$.
In the low temperature regime $\bt\gg1$ the dominant contribution
to the integral
\begin{equation}
\begin{aligned}
 \bra Z(\bt)\ket=N\int_{E_0}^{-E_0}dE \rho(E)e^{-\bt E}
\end{aligned} 
\label{eq:z-int}
\end{equation}
comes from the edge of the spectrum near the ground state $E=E_0$.
Introducing the excitation energy $\cE$
from the ground state,
\begin{equation}
\begin{aligned}
 \cE= E-E_0,
\end{aligned} 
\end{equation}
the eigenvalue density \eqref{eq:wigner} near the ground state $\cE\approx 0$ behaves as
\begin{equation}
\begin{aligned}
 \rho(E)\sim \rt{\cE}.
\end{aligned} 
\label{eq:rho-edge}
\end{equation}
Then \eqref{eq:z-int} is approximated as
\begin{equation}
\begin{aligned}
 \bra Z(\bt)\ket
\approx Ne^{-\bt E_0}\int d\cE \rt{\cE}e^{-\bt\cE}\approx
\frac{N}{\bt^{3/2}}e^{-\bt E_0},
\end{aligned}
\label{eq:edge-z} 
\end{equation}
which reproduces the scaling behavior in \eqref{eq:z-eff}
for $n=1$.
For $n\geq2$, we could not find a simple scaling argument 
to derive \eqref{eq:z-eff} from the integral representation of
$\bra Z(\bt)^n\ket_c$ in \eqref{eq:rho-c}.
It would be interesting to understand the behavior \eqref{eq:z-eff}
directly from \eqref{eq:rho-c}.

\subsubsection{Comparison with the spectral form factor}
The spectral form factor (SFF), defined by
$\bra Z(\ri t)Z(-\ri t)\ket$, is widely studied as a useful diagnostic of quantum
chaos \cite{jost}.
In SFF, 
the exchange of dominance between the connected and the disconnected
parts also occurs at the so-called dip time $t_{\text{dip}}$, 
which can be estimated as follows \cite{Cotler:2016fpe}.
The disconnected part of SFF behaves as
\begin{equation}
\begin{aligned}
 \bra Z(\ri t)\ket\bra Z(-\ri t)\ket
\sim\left(\frac{N}{t^{3/2}}\right)^2,
\end{aligned}
\label{eq:slope} 
\end{equation}
which describes the early time decay of SFF, the so-called ``slope.''
The origin of the factor $1/t^{3/2}$ in \eqref{eq:slope} 
is essentially the same as the calculation
in \eqref{eq:edge-z}.
On the other hand, the connected part grows linearly in time,
which is known as the ``ramp'':
\begin{equation}
\begin{aligned}
 \bra Z(\ri t)Z(-\ri t)\ket_c
\sim t.
\end{aligned} 
\label{eq:ramp} 
\end{equation}
Equating \eqref{eq:slope} and \eqref{eq:ramp}
we find the time scale of ``dip''
\begin{equation}
\begin{aligned}
 t_{\text{dip}}\sim N^{1/2}.
\end{aligned} 
\label{eq:dip}
\end{equation}
The $N$-dependence of $t_{\text{dip}}$ is different from that of the transition temperature
$\bt_{\text{RSB}}$ in \eqref{eq:bt-glass}.
This
is not a contradiction since $\bra Z(\bt)^2\ket$
and the SFF are sensitive to the different aspects
of the two-point correlation $\rho^{(2)}(E_1,E_2)$ of energy eigenvalues:
\begin{equation}
\begin{aligned}
 \bra Z(\bt)^2\ket&=\int dE_1dE_2\,e^{-\bt(E_1+E_2)}\det\Bigl[K(E_i,E_j)\Bigr]_{i,j=1,2},\\
\bra Z(\ri t)Z(-\ri t)\ket&=\int dE_1dE_2\,e^{\ri t(E_1-E_2)}\det\Bigl[K(E_i,E_j)\Bigr]_{i,j=1,2}.
\end{aligned} 
\end{equation}
Namely,  $\bra Z(\bt)^2\ket$ is the Laplace transform of $\rho^{(2)}$
with respect to the total energy, while SFF is the Fourier
transform of $\rho^{(2)}$
with respect to the energy difference.

\subsection{Landscape of free energy}
\label{sec:glass}
The behavior of $\bra Z(\bt)^n\ket_c$ in \eqref{eq:z-eff}
can be 
easily generalized to arbitrary 
term $\cZ_Y(\bt)$ in the expansion \eqref{eq:Y-sum}
\begin{equation}
\begin{aligned}
 \cZ_Y(\bt)\sim e^{-n\bt E_0}\prod_{j=1}^n g_{\text{eff}}^{(j-2)k_j}.
\end{aligned} 
\label{eq:zY-eff}
\end{equation}
When $g_{\text{eff}}\sim 1$,
not only the totally connected and disconnected terms with $Y=[n]$ and 
$Y=[1^n]$, but all terms in the expansion \eqref{eq:Y-sum}
become comparable.

\begin{figure}[thb]
\centering
\subcaptionbox{$|Y|=3$\label{sfig:FY3}}{\includegraphics[width=7.5cm]{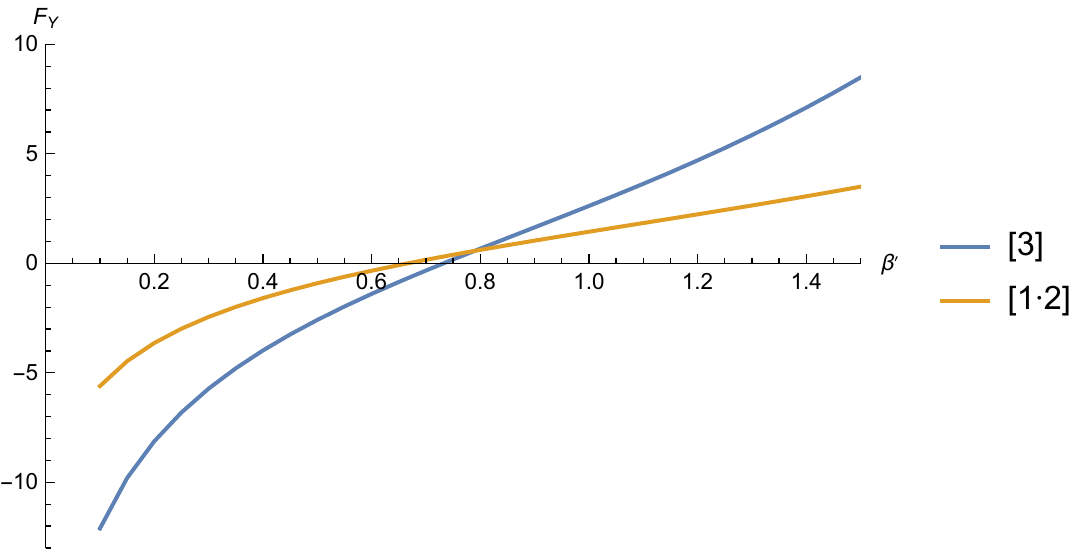}}
\subcaptionbox{$|Y|=4$\label{sfig:FY4}}{\includegraphics[width=7.5cm]{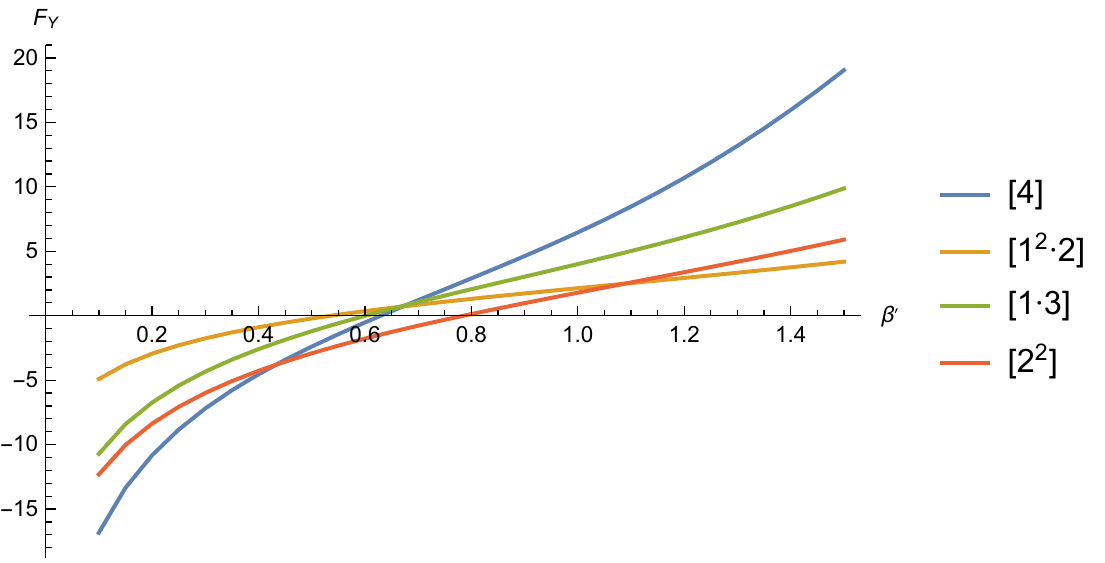}}
  \caption{Plot of $F_Y(\bt)$
for \subref{sfig:FY3} $|Y|=3$ and \subref{sfig:FY4} $|Y|=4$ with $N=60$.
Note that the horizontal axis is $\bt'=\bt/N^{2/3}$.
}
  \label{fig:FY}
\end{figure}

As we did in section \ref{sec:num},
we can numerically study the behavior of $\cZ_Y(\bt)$
using the exact result in section \ref{sec:exact}.
Let us define the free energy of $\cZ_Y(\bt)$
relative to the disconnected part
\begin{equation}
\begin{aligned}
 F_Y(\bt)=\log\cZ_Y(\bt)-\log \bra Z(\bt)\ket^n.
\end{aligned} 
\label{eq:FY-def}
\end{equation}
In Figure~\ref{fig:FY}, we plot $F_Y(\bt)$ for various $Y$'s
with $|Y|=3$ and $|Y|=4$ for $N=60$. 
We have checked numerically
that the plot for $N=100$ 
is almost identical to the case of $N=60$ in Figure \ref{fig:FY}
in the regime $\bt'\lesssim1.5$.
As one can see from Figure~\ref{fig:FY},
all $F_Y(\bt)$ become almost degenerate around the transition point $\bt'\sim 1$.
For $|Y|=4$ in Figure~\ref{sfig:FY4}, the contribution of $Y=[2^2]$ is relatively small
compared to the other partitions, but $F_{[2^2]}$ is still of order $\cO(N^0)$
and it is not suppressed by a negative power of $N$.
 
Near the transition around $T=T_{\text{RSB}}$,
all terms $\cZ_Y(\bt)$ in the expansion \eqref{eq:Y-sum}
contribute with almost equal weight.
This transition is not a true phase transition in the sense of Ehrenfest classification:
it does not involve any discontinuity in the derivative of free energy.
This is in contrast to the confinement/deconfinement transition in 4d $\mathcal{N}=4$
SYM 
and its dual Hawking-Page transition \cite{Witten:1998zw,Hawking:1982dh}, 
which is very likely a first order phase transition \cite{Aharony:2005bq,Aharony:2003sx}.
In that case the free energy changes from $\cO(N^0)$ to $\cO(N^2)$
at the transition, but in our case there is no
such large change of free energy: the free energy
stays the same order below and above the transition. 
In fact, the exponential factor $e^{-n\bt E_0}$
of $\cZ_Y$ in \eqref{eq:zY-eff} is common for all $Y$.
Moreover, the number of terms (or the number of partitions $p(n)$) 
in the expansion \eqref{eq:Y-sum}
grows rapidly as $n$ increases
\begin{equation}
\begin{aligned}
 p(n)\sim \exp\Biggl(4\pi\rt{\frac{n}{24}}\Biggr),\qquad (n\gg1).
\end{aligned} 
\label{eq:pn}
\end{equation}
Thus, each term in the expansion \eqref{eq:zY-eff}
can be thought of as a 
local minimum of free energy landscape
labeled by a Young diagram $Y$,
and the number of local minima becomes very large \eqref{eq:pn}
even for relatively small $n$ (e.g. $p(10)=42,~p(25)=1958$).  

When the temperature is below the RSB transition  $T<T_{\text{RSB}}$,
the totally connected contribution $Y=[n]$
becomes dominant.
One can see this explicitly for $n=3,4$ in Figure~\ref{fig:FY}:
the blue curve labeled by $Y=[n]$ becomes the largest contribution
in the low temperature regime $\bt'\gtrsim 1$.
Thus we confirmed numerically the behavior claimed in \eqref{eq:RSB}.

\subsection{Low temperature regime}
\label{sec:lowT}
\begin{figure}[thb]
\centering
\subcaptionbox{$r_2(\bt)$\label{sfig:r2}}{\includegraphics[width=4.5cm]{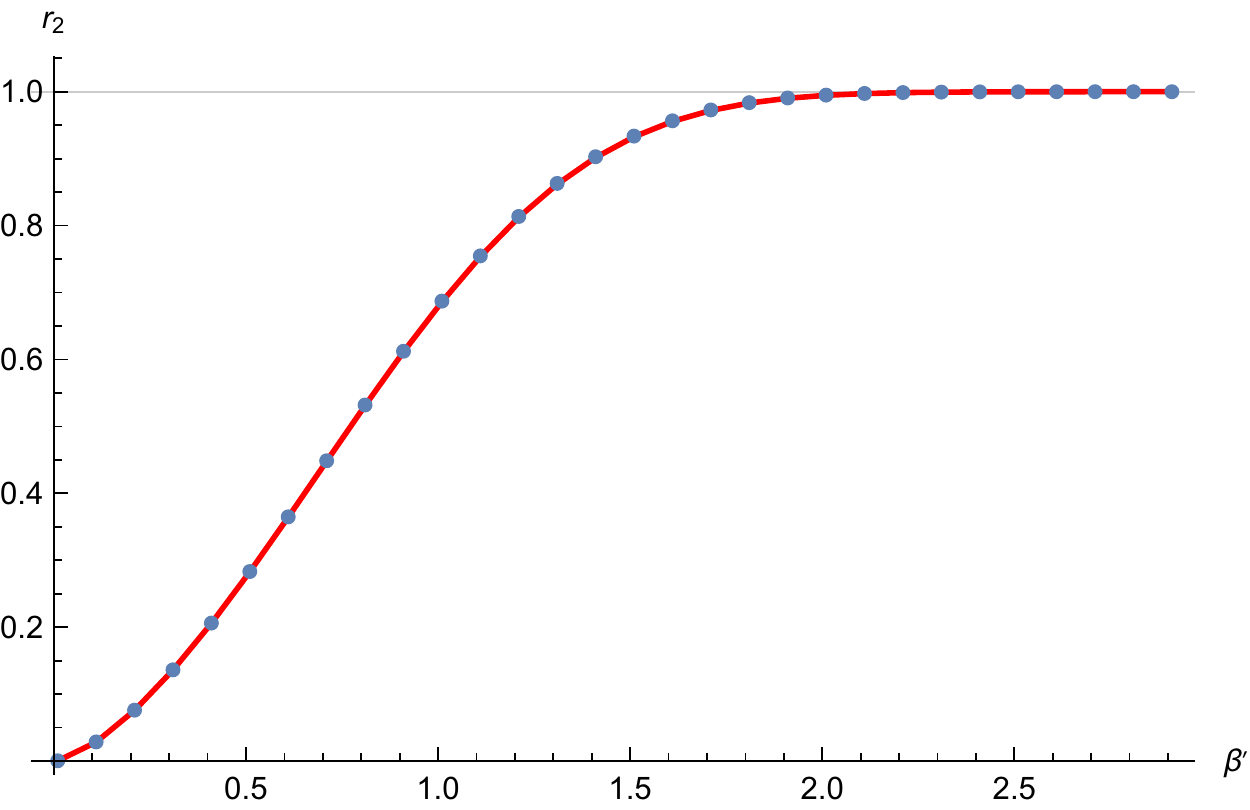}}
\hskip2mm
\subcaptionbox{$r_3(\bt)$\label{sfig:r3}}{\includegraphics[width=4.5cm]{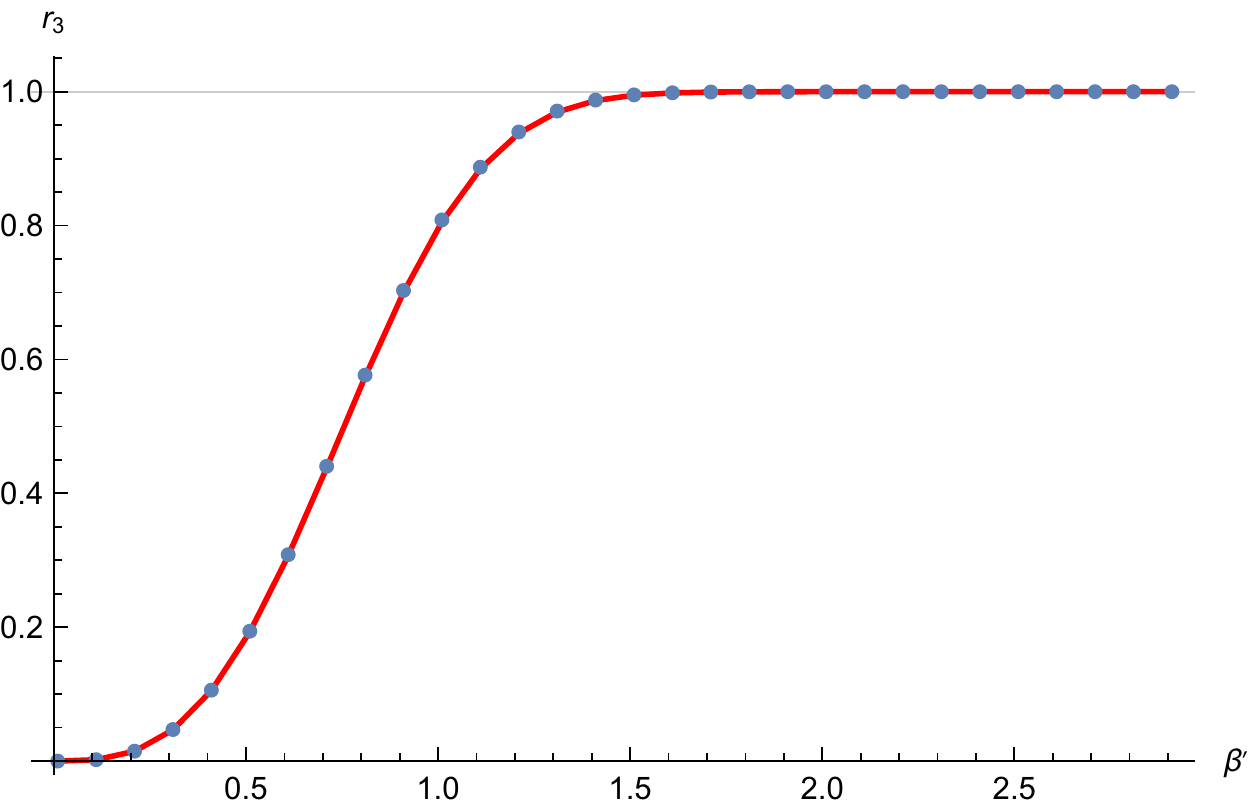}}
\hskip2mm
\subcaptionbox{$r_4(\bt)$\label{sfig:r4}}{\includegraphics[width=4.5cm]{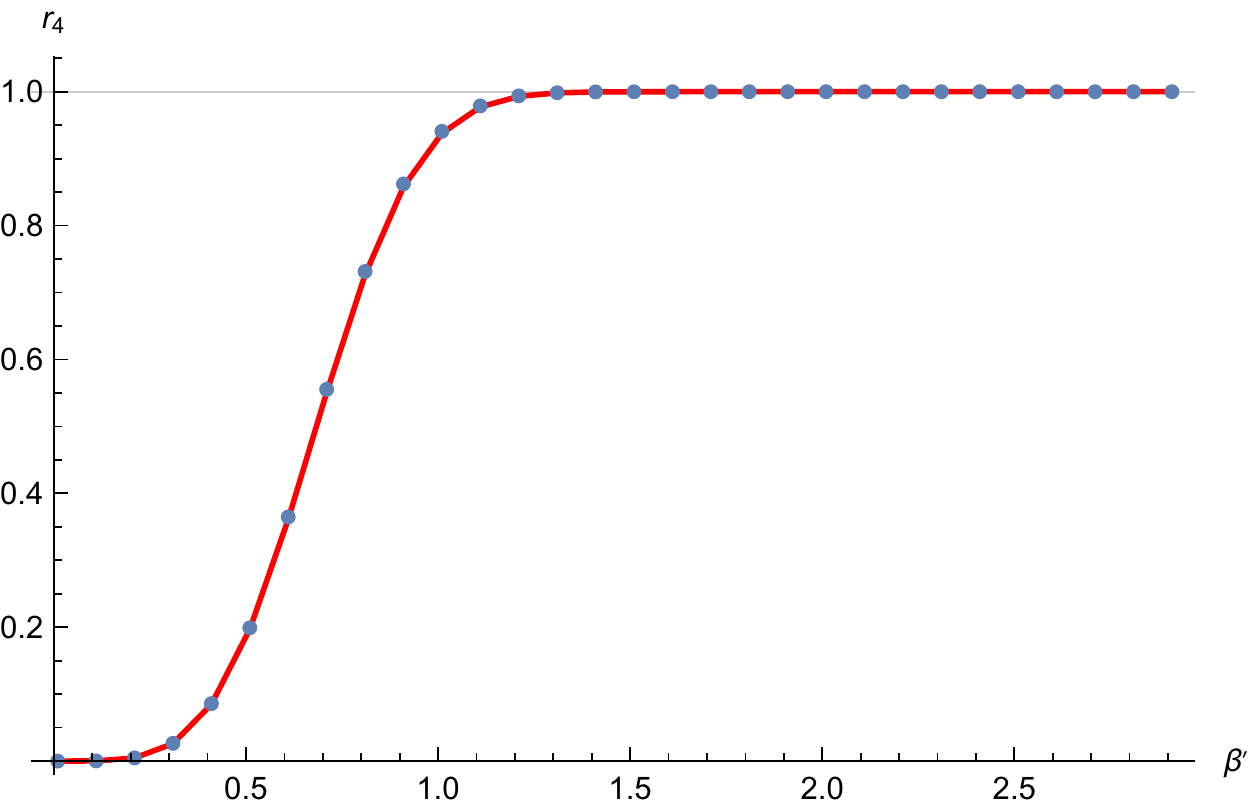}}
  \caption{Plot of $r_n(\bt)$
for \subref{sfig:r2} $n=2$, \subref{sfig:r3} $n=3$, and \subref{sfig:r4} $n=4$.
Note that the horizontal axis is $\bt'=\bt/N^{2/3}$.
The red curve is the plot for $N=60$, while the blue dots represent 
the value of $r_n(\bt)$ for $N=100$.
}
  \label{fig:ratio}
\end{figure}

In the previous subsection \ref{sec:glass} we have seen that
the totally connected part $\bra Z(\bt)^n\ket_c$ becomes dominant in
the low temperature regime $T< T_{\text{RSB}}$.
We find that in this regime $\bra Z(\bt)^n\ket_c$
is further approximated by the 
first term in \eqref{eq:z2-A} and \eqref{eq:z34-A} 
\begin{equation}
\begin{aligned}
 \bra Z(\bt)^n\ket_c \approx \Tr A(n\bt)=\bra Z(n\bt)\ket,\quad (T < T_{\text{RSB}}).
\end{aligned} 
\label{eq:zc-approx}
\end{equation}
To see this, let us consider the ratio $r_n(\bt)$ 
between the both sides of \eqref{eq:zc-approx}
\begin{equation}
\begin{aligned}
 r_n(\bt)=\frac{\bra Z(\bt)^n\ket_c}{\bra Z(n\bt)\ket}.
\end{aligned} 
\label{eq:rn-def}
\end{equation}
In Figure \ref{fig:ratio}, we plot  $r_n(\bt)$
for $n=2,3,4$ as a function of $\bt'=\bt/N^{2/3}$
for $N=60$ (red curve) and $N=100$ (blue dots).
As we can see from Figure \ref{fig:ratio},
$r_n(\bt)$ approaches $1$ as $\bt'$ increases,
which confirms the claim in \eqref{eq:zc-approx}.
Finally, combining \eqref{eq:zc-approx} and  
\eqref{eq:RSB}, in the low temperature regime
we find
\begin{equation}
\begin{aligned}
 \bra Z(\bt)^n\ket\approx \bra Z(\bt)^n\ket_c
\approx \bra Z(n\bt)\ket,\qquad(T< T_{\text{RSB}}).
\end{aligned} 
\label{eq:lowT}
\end{equation}

\section{Possible implications for  multiply-coupled SYK models}
\label{sec:SYK}

In this section, we will discuss possible implications
of our findings in the GUE random matrix model
for multiply-coupled SYK models and their dual JT gravity.

The SYK model exhibits a chaotic behavior
which is diagnosed by the SFF \cite{Cotler:2016fpe} as well as
the out-of-time ordered correlators (OTOCs) \cite{kitaev,Maldacena:2016hyu,Kitaev:2017awl}.
The Lyapunov exponent extracted from the OTOC of SYK model
saturates the chaos bound \cite{bound} which is consistent with
the holographically dual black hole interpretation.

The level statistics of energy eigenvalues of the SYK model is well approximated by
the random matrix models \cite{Garcia-Garcia:2016mno,Cotler:2016fpe},
which is also a characteristic feature of the quantum chaos.
At low energy, the SYK model is described by
the Schwarzian theory  whose density of states is given by \cite{Garcia-Garcia:2017pzl,Stanford:2017thb}
\begin{equation}
\begin{aligned}
 \rho_{\text{Sch}}(\cE)=\sinh(2\pi\rt{\cE}).
\end{aligned} 
\label{eq:rho-sch}
\end{equation} 
Although this is different from
the simple semi-circle law in \eqref{eq:wigner},
$\rho_{\text{Sch}}(\cE)$ exhibits the 
same square-root behavior $\rho(\cE)\sim\rt{\cE}$ \eqref{eq:rho-edge}
near the edge of the spectrum $\cE=0$.

The above discussion suggests that the RSB transition observed in the GUE
random matrix model in the previous section \ref{sec:RSB}
also appears in the multiply-coupled SYK models, 
where the expectation value of $\bra Z(\bt)^n\ket$ is taken by the disorder average
over the random coupling in the Hamiltonian of SYK model.
Recall that the scaling of the effective string coupling $g_{\text{eff}}$
in \eqref{eq:g-eff} is sensitive only to the behavior
of eigenvalue density near the edge of the spectrum.
Thus the behavior of $g_{\text{eff}}$ is expected to be the same
for both the GUE random matrix model and the Schwarzian theory,
and the transition occurs when $g_{\text{eff}}\sim 1$.

However, the RSB and the role of replica non-diagonal saddle point
in the SYK model are still controversial
\cite{Gur-Ari:2018okm,Wang:2018ijz,Saad:2018bqo,Arefeva:2018vfp};
It is argued in \cite{Gur-Ari:2018okm}
that RSB does not occur in the SYK model.
In \cite{Wang:2018ijz,Saad:2018bqo,Arefeva:2018vfp}
the replica non-diagonal saddle points are found by numerical analysis
but it is argued that such saddle points are always sub-dominant
compared to the diagonal one.\footnote{See also \cite{Arefeva:2019wjb}
for the RSB in a zero-dimensional version of the SYK model.}
In a recent paper \cite{SSS},
it is argued that 
a non-perturbative completion
of Schwarzian theory and its dual JT gravity
is given by a certain double-scaled matrix model,
and it is suggested that the replica non-diagonal saddle point
does play an essential role.

Interestingly, it is shown in \cite{Qi}
that by turning on a mass term connecting the fermions
living on different replicas,
the partition function of
coupled SYM models exhibits a first order
phase transition, which is very similar
to the transition we observed for $\bra Z(\bt)^2\ket$
in the GUE random matrix model 
(see also \cite{Kim:2019upg}
for the symmetry breaking in coupled SYK models).

In the rest of this section, we will have in mind
a certain soft breaking of replica symmetry as in \cite{Qi},
and see what kind of dual gravity picture is expected 
based on the observation in the previous section  \ref{sec:RSB}.
Our argument is merely a heuristic one and 
we have not performed an actual calculation of
the multiply-coupled SYK models with $n\ge3$ replicas generalizing \cite{Qi}.
It would be very interesting to study such multiply-coupled SYK models.

Based on the fact that the level statistics of the SYK model
is described by a random matrix model,
we expect that the phase structure
of the GUE random matrix model we observed in the previous 
section captures the essential part of the low energy dynamics
of multiply-coupled SYK models.

\subsection{Wormhole networks}

The bulk gravity picture of $n=2$ coupled SYK models
in \cite{Qi}
is that the two terms in the expansion
of $\bra Z(\bt)^2\ket$ in \eqref{eq:zconn-exp}
correspond to the two geometries with different topology:
the first term $\bra Z(\bt)\ket^2$ corresponds to the two disconnected disks
associated with the two Euclidean black holes,
while the second term 
$\bra Z(\bt)^2\ket_c$ corresponds to 
the Euclidean wormhole connecting the two boundaries
with the topology of annulus.
This is depicted schematically as:
\begin{align}
\begin{aligned}
\bra Z(\bt)^2\ket&=\bra Z(\bt)\ket^2\quad+\quad\bra Z(\bt)^2\ket_c\\
&=
\tikz[baseline=0pt]{
\filldraw[fill=orange] (0,0) ellipse (0.08 and 0.3);
\filldraw[fill=orange] (1,0) ellipse (0.08 and 0.3);
}
\hskip5mm
+
\hskip5mm
\tikz[baseline=0pt]{
\filldraw[fill=orange] (0,0) ellipse (0.08 and 0.3);
\filldraw[fill=orange] (1,0) ellipse (0.08 and 0.3);
\draw[ bend right = 30] (0,0.3) to (1,0.3);
\draw[ bend left = 30] (0,-0.3) to (1,-0.3);
}\quad.
\end{aligned} 
\label{eq:wormhole}
\end{align}

This picture for $n=2$ replicas
suggests that the holographic dual geometry of the connected correlator
is literally connected by an Euclidean wormhole,
in the same spirit as the ER=EPR conjecture \cite{Maldacena:2013xja}.
Assuming this interpretation, we can draw a cartoon
of spacetime geometry for all terms $\cZ_Y(\bt)$
in the expansion \eqref{eq:Y-sum}.
For $n=3$ we have three terms $Y=[1^3],[1\cdot2],[3]$
in the expansion \eqref{eq:zconn-exp},
where the corresponding
spacetime geometries look like
\begin{equation}
\begin{aligned}
  \cZ_{[1^3]}(\bt)=\bra Z(\bt)\ket^3=\quad
\tikz[baseline=0pt]{
\filldraw[fill=orange] (0,0.5) ellipse (0.3 and 0.08);
\filldraw[fill=orange] (-0.433,-0.25) ellipse [x radius=0.3, y radius=0.08, rotate=-60];
\filldraw[fill=orange] (0.433,-0.25) ellipse [x radius=0.3, y radius=0.08, rotate=60];
}\quad,
\end{aligned} 
\end{equation}
\begin{equation}
\begin{aligned}
 \cZ_{[1\cdot2]}(\bt)=3\bra Z(\bt)\ket\bra Z(\bt)^2\ket_c=\quad
\tikz[baseline=0pt]{
\filldraw[fill=orange] (0,0.5) ellipse (0.3 and 0.08);
\filldraw[fill=orange] (-0.433,-0.25) ellipse [x radius=0.3, y radius=0.08, rotate=-60];
\filldraw[fill=orange] (0.433,-0.25) ellipse [x radius=0.3, y radius=0.08, rotate=60];
\draw[ bend right = 30] (0.3,0.5) to (0.583,0.0098);
\draw[ bend right = 20] (-0.3,0.5) to (0.283,-0.5098);
}
\quad
+\quad
\tikz[baseline=0pt]{
\filldraw[fill=orange] (0,0.5) ellipse (0.3 and 0.08);
\filldraw[fill=orange] (-0.433,-0.25) ellipse [x radius=0.3, y radius=0.08, rotate=-60];
\filldraw[fill=orange] (0.433,-0.25) ellipse [x radius=0.3, y radius=0.08, rotate=60];
\draw[ bend left = 30] (-0.3,0.5) to (-0.583,0.0098);
\draw[ bend left = 20] (0.3,0.5) to (-0.283,-0.5098);
}
\quad+\quad
\tikz[baseline=0pt]{
\filldraw[fill=orange] (0,0.5) ellipse (0.3 and 0.08);
\filldraw[fill=orange] (-0.433,-0.25) ellipse [x radius=0.3, y radius=0.08, rotate=-60];
\filldraw[fill=orange] (0.433,-0.25) ellipse [x radius=0.3, y radius=0.08, rotate=60];
\draw[ bend right = 20] (0.583,0.0098) to (-0.583,0.0098);
\draw[ bend right = 30] (0.283,-0.5098) to (-0.283,-0.5098);
}\quad,
\end{aligned} 
\label{eq:pic-12}
\end{equation}
and
\begin{equation}
\begin{aligned}
 \cZ_{[3]}(\bt)=\bra Z(\bt)^3\ket_c=\quad
\tikz[baseline=0pt]{
\filldraw[fill=orange] (0,0.5) ellipse (0.3 and 0.08);
\filldraw[fill=orange] (-0.433,-0.25) ellipse [x radius=0.3, y radius=0.08, rotate=-60];
\filldraw[fill=orange] (0.433,-0.25) ellipse [x radius=0.3, y radius=0.08, rotate=60];
\draw[ bend right = 30] (0.3,0.5) to (0.583,0.0098);
\draw[ bend right = 30] (0.283,-0.5098) to (-0.283,-0.5098);
\draw[ bend right = 30] (-0.583,0.0098) to (-0.3,0.5);
}\quad.
\end{aligned} 
\label{eq:pic-3}
\end{equation}
Note that the coefficient $3$ in $\cZ_{[1\cdot2]}(\bt)=3\bra Z(\bt)\ket\bra Z(\bt)^2\ket_c$
has a natural interpretation as the number of ways to connect
two boundaries by a wormhole out of three boundaries, 
as shown in \eqref{eq:pic-12}.
One can also check that the coefficient appearing in the expansion of
$\bra Z(\bt)^4\ket$ in \eqref{eq:zconn-exp}
has a similar interpretation.

For general $n$,
from the result of random matrix model in \eqref{eq:RSB},
we can draw the following dual spacetime picture:
at high temperature $T>T_{\text{RSB}}$ the
bulk geometry is $n$ disconnected disks
representing $n$ disconnected Euclidean black holes, while
at low temperature $T<T_{\text{RSB}}$ 
the corresponding bulk geometry is an $n$-pronged wormhole
connecting the $n$ boundaries (see \eqref{eq:pic-3} for the example
of three-pronged wormhole). 
Around the RSB transition temperature $T\sim T_{\text{RSB}}$,
we speculate that the dual gravity side
does not correspond to a classical geometry, but 
it is described by 
a superposition of random wormhole networks.
It would be very interesting to understand the
implication of this RSB transition in the bulk gravity theory.

\subsection{Thermofield $n$-tuple and ground state}

In the previous section \ref{sec:lowT}, we observed that
$\bra Z(\bt)^n\ket$ is reduced to $\bra Z(\bt)^n\ket_c\approx
\bra Z(n\bt)\ket$
in the low temperature regime \eqref{eq:lowT}.
Let us consider a possible interpretation of this behavior.
We first notice that our starting point
$Z(\bt)^n$ is written as a trace
over the $n$-th power of the Hilbert space
$\cH$ of a single system
\begin{equation}
\begin{aligned}
 Z(\bt)^n=\Tr_{\cH^{\otimes n}}\underbrace{e^{-\bt H}\otimes \cdots \otimes
e^{-\bt H}}_n~.
\end{aligned} 
\end{equation}
Since $Z(\bt)^n$
is defined as a trace in $\cH^{\otimes n}$,
it is natural to express its low temperature
limit $\bra Z(n\bt)\ket$ in \eqref{eq:lowT} as a quantity involving
$\cH^{\otimes n}$.
However, this seems impossible since $Z(n\bt)=\Tr_{\cH}e^{-n\bt H}$
is a trace on the single Hilbert space $\cH$.
This tension is resolved by assuming that the low temperature state in $\cH^{\otimes n}$
is highly entangled.
We propose that $\bra Z(n\bt)\ket$ should be written as
\begin{equation}
\begin{aligned}
 \bra Z(n\bt)\ket  =\bra \text{TF}_n|\text{TF}_n\ket,
\end{aligned} 
\label{eq:z-TFn}
\end{equation}
where $|\text{TF}_n\ket$ 
denotes the ``thermofield $n$-tuple state''
\begin{equation}
\begin{aligned}
 |\text{TF}_n\ket =\sum_{i}e^{-\frac{n\bt}{2}E_i}
\underbrace{|i\ket\otimes\cdots\otimes |i\ket}_n~~\in\mathcal{H}^{\otimes n}.
\end{aligned} 
\label{eq:TF-n}
\end{equation}
This is a generalization of the thermofield double state $|\text{TFD}\ket$
for $n=2$
\begin{equation}
\begin{aligned}
 |\text{TFD}\ket=|\text{TF}_2\ket.
\end{aligned} 
\end{equation}

In the low temperature limit, the partition function is approximated by
the expectation value in the ground state.
From \eqref{eq:lowT} and \eqref{eq:z-TFn},
we expect that the ground state $|\Psi_\text{gnd}\ket$ of multiply-coupled system is 
approximated by the thermofield $n$-tuple state
\begin{equation}
\begin{aligned}
 |\Psi_\text{gnd}\ket\approx |\text{TF}_n\ket.
\end{aligned} 
\end{equation}
This property is indeed observed for $n=2$ in \cite{Qi}.

Our relations \eqref{eq:lowT} and \eqref{eq:z-TFn}
are also consistent with the picture
of spacetime geometry built by entanglement
\cite{VanRaamsdonk:2010pw} and the ER=EPR conjecture
\cite{Maldacena:2013xja}. 
In fact, the relation between 
the thermofield double state $|\text{TFD}\ket$
and the wormhole picture of $\bra Z(\bt)^2\ket_c$
depicted in \eqref{eq:wormhole}
is the motivation for the proposal in
\cite{VanRaamsdonk:2010pw,Maldacena:2013xja}.
Then it is natural to generalize this picture
to the case of multiple boundaries.
For instance,
the thermofield triple state $|\text{TF}_3\ket$
corresponds to
the three-pronged wormhole depicted in 
\eqref{eq:pic-3}.\footnote{In our Universe,
it is actually observed that
there exist ternary systems of supermassive black holes \cite{ternary}.}
We emphasize that only the totally connected part is dominant
in the low temperature limit, and hence the ground state
$|\Psi_\text{gnd}\ket\approx |\text{TF}_n\ket$
would be dual to an $n$-pronged wormhole.
In the spirit of \cite{VanRaamsdonk:2010pw,Maldacena:2013xja},
this $n$-pronged wormhole can be thought of as a geometric
manifestation of the entanglement in the state \eqref{eq:TF-n}.

\acknowledgments
This work was supported in part by JSPS KAKENHI Grant Number 16K05316.


\end{document}